\documentclass[lettersize,journal]{IEEEtran}
\usepackage{amsmath,amsfonts, amssymb}
\usepackage{array}
\usepackage[caption=false,font=normalsize,labelfont=sf,textfont=sf]{subfig}
\usepackage{textcomp}
\usepackage{stfloats}
\usepackage{url}
\usepackage{verbatim}
\usepackage{graphicx}
\usepackage{tabularx}
\usepackage{booktabs}
\usepackage{algorithm}
\usepackage[noend]{algpseudocode}
\usepackage{multirow} 
\usepackage{makecell} 
\usepackage{lipsum}
\usepackage{scalerel}
\usepackage{comment}
\usepackage{enumitem}
\setlist[itemize]{noitemsep, nolistsep}
\setlist[enumerate]{noitemsep, nolistsep}
\usepackage{xcolor}

\def\BibTeX{{\rm B\kern-.05em{\sc i\kern-.025em b}\kern-.08em
    T\kern-.1667em\lower.7ex\hbox{E}\kern-.125emX}}
\usepackage{balance}

\usepackage{pifont}
\newcommand{\circled}[1]{\textcircled{\scriptsize #1}}

\usepackage[utf8]{inputenc}
\usepackage{kotex}

\usepackage{xspace}
\newcommand{\Ours}{\textsc{Cosmos}\xspace}

\begin{document}

\title{\Ours: A CXL-Based Full In-Memory System for Approximate Nearest Neighbor Search}

\author{Seoyoung~Ko,~Hyunjeong~Shim,~Wanju~Doh,~Sungmin~Yun,~Jinin~So,~Yongsuk~Kwon,~Sang-Soo~Park, Si-Dong~Roh,~Minyong~Yoon,~Taeksang~Song,~and~Jung~Ho~Ahn,~\IEEEmembership{Senior Member, IEEE}

\thanks{This work was partly supported by Samsung Electronics Co., Ltd (IO250301-12185-01) and IITP (RS-2021-II211343 and RS-2023-00256081).}
\thanks{Seoyoung Ko, Hyunjeong Shim, Wanju Doh, Sungmin Yun, and Jung Ho Ahn are with Seoul National University, Seoul 08826, South Korea. E-mail: \{seoyoungko, simhj1212, wj.doh, sungmin.yun, gajh\}@snu.ac.kr.}
\thanks{Jinin So, Yongsuk Kwon, Sang-Soo Park, Si-Dong Roh, Minyong Yoon, and Taeksang Song are with Samsung Electronics Corporation, Hwaseong-si, Gyeonggi-do 18448, South Korea. E-mail: \{jinin.so, yssh.kwon, ss23.park, sidong.roh, casper.yoon, taeksang.song\}@samsung.com.}
}



\maketitle

\begin{abstract}
Retrieval-Augmented Generation (RAG) is crucial for improving the quality of large language models by injecting proper contexts extracted from external sources.
RAG requires high-throughput, low-latency Approximate Nearest Neighbor Search (ANNS) over billion-scale vector databases. 
Conventional DRAM/SSD solutions face capacity/latency limits, whereas specialized hardware or RDMA clusters lack flexibility or incur network overhead.
We present \Ours, integrating general-purpose cores within CXL memory devices for full ANNS offload and introducing rank-level parallel distance computation to maximize memory bandwidth.
We also propose an adjacency-aware data placement that balances search loads across CXL devices based on inter-cluster proximity.
Evaluations on SIFT1B and DEEP1B traces show that \Ours achieves up to 6.72$\times$ higher throughput than the baseline CXL system and 2.35$\times$ over a state-of-the-art CXL-based solution, demonstrating scalability for RAG pipelines.
\end{abstract}

 \begin{IEEEkeywords}
 CXL, Approximate Nearest Neighbor Search, Processing Near Memory, Retrieval-Augmented Generation
 \end{IEEEkeywords}

\section{Introduction}

\IEEEPARstart{R}{etrieval-Augmented}
 Generation (RAG) enhances Large Language Models (LLMs) by dynamically retrieving relevant information from external databases, enabling more accurate and contextually appropriate responses~\cite{arxiv-2024-RAG-tradeoff}.
Techniques such as Agentic RAG~\cite{arxiv-2025-AgenticRAG} further improve quality through iterative retrieval.
Central to RAG is Approximate Nearest Neighbor Search (ANNS), which enables fast retrieval in high-dimensional vector spaces by efficiently identifying the top-k most relevant vectors---those closest to a given query---based on a similarity metric.
As datasets grow, efficient and scalable ANNS is critical for real-time inference.

Handling billion-scale ANNS workloads challenges traditional systems.
DRAM lacks capacity, whereas SSDs suffer from high latency incompatible with fine-grained ANNS access patterns~\cite{atc-2024-SmartANN, arxiv-2024-second-tier-vectorDB}.
Alternative solutions such as conventional Processing Near Memory (PNM)~\cite{atc-2024-SmartANN, isca-2024-NDSearch} lack flexibility, and RDMA clusters~\cite{arxiv-2024-second-tier-vectorDB} incur network latency and complexity.

Compute Express Link (CXL) offers a promising solution with high-bandwidth, low-latency memory expansion~\cite{atc-2022-directCXL}, beneficial for latency-sensitive RAG~\cite{arxiv-2024-RAG-tradeoff}.
Still, ANNS remains memory-bandwidth bound, dominated by distance calculations~\cite{nips-2019-diskann}.
Distributing indices across multiple CXL devices requires intelligent data placement to avoid load imbalance.

In this paper, we propose \Ours, a full in-memory ANNS system using compute-capable CXL devices. \Ours makes the following three key contributions:
\begin{itemize}[leftmargin=*]
  \item \textbf{Full ANNS Offload via CXL GPCs:} \Ours integrates programmable general-purpose cores (GPCs) in CXL memory controllers for local ANNS execution, eliminating host intervention and PCIe traffic during search.
  \item \textbf{Rank-level Parallel Distance Computation:} \Ours exploits DRAM rank-level parallelism for concurrent distance computation, reducing data movement and maximizing memory bandwidth.
  \item \textbf{Adjacency-aware Data Placement:} 
  A lightweight algorithm uses cluster metadata to distribute vectors across CXL devices, balancing load and enabling parallelism without runtime profiling.
\end{itemize}

\section{Background}

\begin{figure}[!t]
  \center
  \includegraphics[width=\columnwidth]{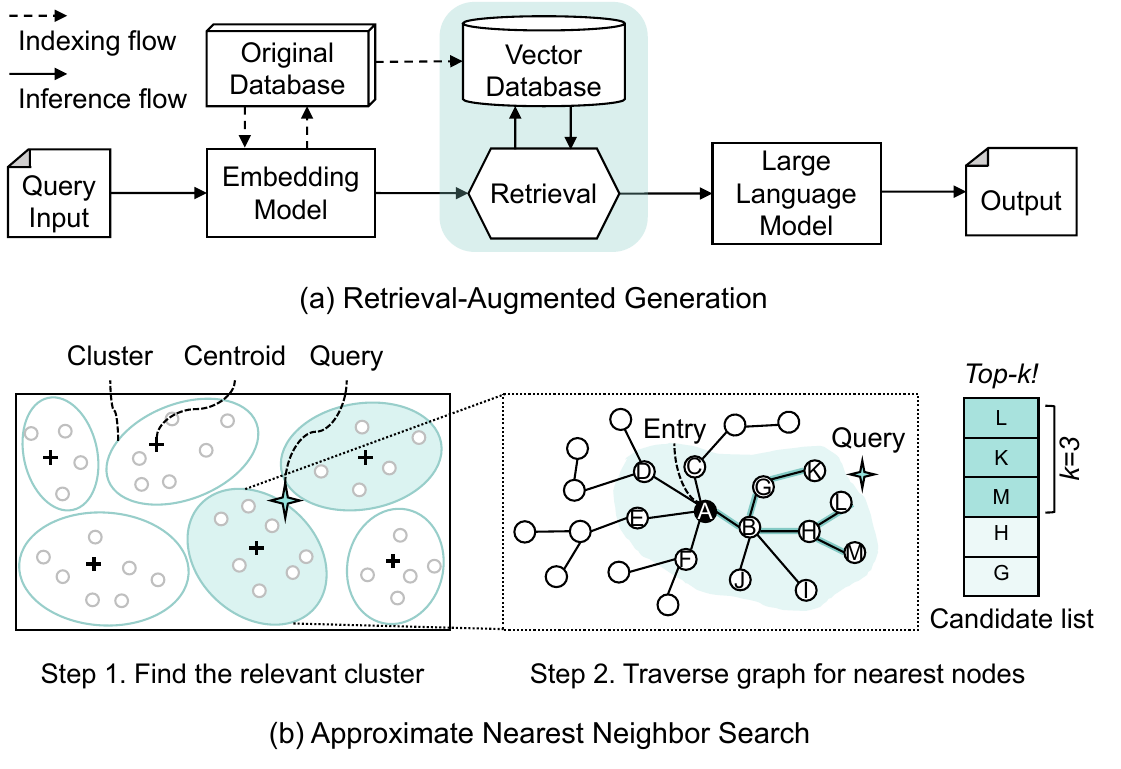}
  \vspace{-0.2in}
  \caption{Overview of retrieval-augmented generation (RAG) and approximate nearest neighbor search (ANNS).}
  \label{fig:background}
   \vspace{-0.06in}
\end{figure}

RAG enhances LLMs by retrieving relevant information from external vector databases during inference (Fig.~\ref{fig:background}(a)).
This typically involves indexing documents into a vector database and, during inference, embedding user queries to retrieve similar documents.
Recently, Agentic RAG~\cite{arxiv-2025-AgenticRAG} improves quality via iterative search strategies.
The core retrieval mechanism relies on finding vectors (documents) most similar to the query vector.

Exact Nearest Neighbor Search (ENNS) guarantees the highest accuracy; however, its linear scaling with dataset size in computational cost makes it impractical for billion-scale data~\cite{atc-2024-SmartANN}.
By contrast, ANNS offers a trade-off between accuracy and efficiency, enabling real-time search. 

ANNS methods include graph-based and cluster-based approaches~\cite{arxiv-2024-second-tier-vectorDB, vldb-24-chameleon}.
The former links similar vectors and searches greedily along edges, but can suffer from irregular memory access.
The latter partitions data, finds the closest cluster(s) to the query, and searches within them, which offers better memory efficiency but suffers from potential read amplification.
Hybrid approaches combine these, restricting graph traversal to relevant clusters, improving efficiency for large datasets while maintaining quality in billion-scale datasets.
Fig.~\ref{fig:background}(b) shows an exemplar hybrid-based ANNS process when k is 3.

\begin{figure}[!t]
  \center
  \includegraphics[width=\columnwidth]{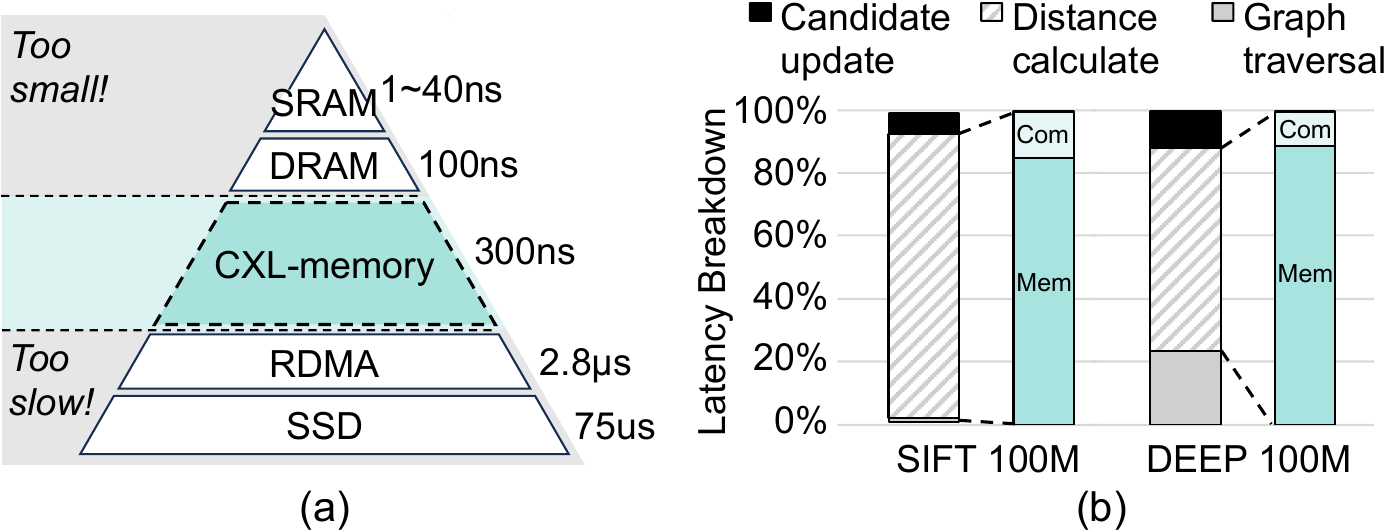}
  \caption{(a) Memory latency hierarchy highlighting the potential of CXL-attached memory as a new tier between DRAM and RDMA/SSD in terms of latency and capacity. (b) Latency breakdown of graph-based ANN search on large-scale datasets (SIFT and DEEP with 100M vectors).}
  \label{fig:motivation}
   \vspace{-6pt}
\end{figure}
\noindent

\section{Motivation}

\subsection{Scalability Challenges in Billion-Scale ANNS}

Billion-scale ANNS demands vast memory (terabytes)~\cite{atc-2023-CXLANNS, atc-2024-SmartANN}, exceeding single-node DRAM limits.
Using SSDs introduces high latency (tens of microseconds), and their coarse-grained access (kilobyte-scale pages) is unsuitable for fine-grained ANNS, potentially dominating search time~\cite{atc-2024-SmartANN, arxiv-2024-second-tier-vectorDB}.

SSD-based PNM accelerators reduce data movement~\cite{atc-2024-SmartANN, isca-2024-NDSearch}, but they lack flexibility against evolving algorithms or parameters.
RDMA-based multi-node clusters offer lower latency than SSDs but still suffer network overhead (few to several microseconds~\cite{arxiv-2024-second-tier-vectorDB}, Fig.~\ref{fig:motivation}(a)) and add complexity.

Compute Express Link (CXL), a PCIe-based interconnect standard, has emerged as a promising alternative. CXL provides direct load/store access to expanded memory with near-native-DRAM latency (few hundred nanoseconds, Fig.~\ref{fig:motivation}(a)) and high bandwidth, eliminating network overhead~\cite{atc-2022-directCXL}. This low latency is critical for RAG, where retrieval time significantly impacts overall performance, especially with iterative techniques like Agentic RAG. Studies show that retrieval accounts for 36\% of the time-to-first-token in a vanilla RAG and up to 97\% in scenarios involving frequent re-retrieval~\cite{arxiv-2024-RAG-tradeoff}. 

\subsection{Leveraging Compute-capable CXL Devices}

\begin{figure*}[!t]
  \center
  \includegraphics[width=0.91\linewidth]{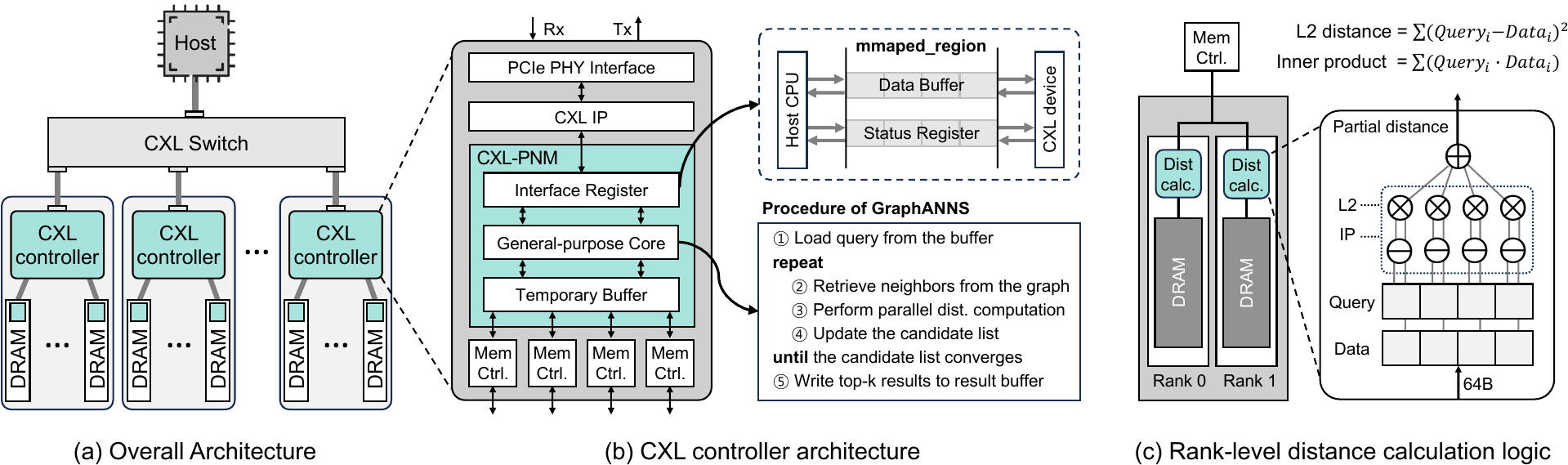}
  \vspace{-0.07in}
  \caption[width=0.99\linewidth]{
  (a) Overview of the system architecture. (b) CXL controller architecture featuring a general-purpose core for executing graph-based ANN search via a memory-mapped host interface. (c) Rank-level distance calculation logic that enables parallel L2 and inner product calculations across memory ranks.
  }
  \label{fig:overview}
  \vspace{-0.06in}
\end{figure*}

Recent CXL memory devices optionally incorporate near-memory compute capabilities~\cite{asplos-2025-pim-is-all-you-need, hpca-2024-cxl-pnm}.
Performing computation near memory reduces data transfer and alleviates bottlenecks for memory-intensive workloads like ANNS.
We propose \emph{offloading the entire ANNS pipeline to compute-capable CXL devices} featuring programmable General-Purpose Cores (GPCs).
GPCs offer flexibility over fixed accelerators, adapting to diverse datasets and parameters (see Table~\ref{tbl:parameter} for the BigANN benchmark~\cite{bigann}) without redesign, which mitigates over/under-provisioning issues on the fixed accelerators.

While offloading helps, ANNS remains bottlenecked by memory bandwidth, primarily due to distance calculations loading large vectors (Fig.~\ref{fig:motivation}(b)).
Our architecture addresses this by integrating a GPC with a DRAM \emph{rank-level processing unit (PU)}.
This exploits DRAM's internal parallelism by partitioning vector dimensions across ranks, computing partial distances concurrently within each rank,
which reduces data movement and improves bandwidth efficiency.  

However, billion-scale datasets require distributing the index across \emph{multiple} CXL devices.
Na\"ive distribution can lead to load imbalance when co-accessed data resides on the same device. 
We address this with an \emph{adjacency-aware cluster placement} algorithm that assigns clusters based on proximity, balancing load and enabling parallel search.

\begin{table}[tb!]
    \centering
    \caption{ANN datasets and search parameters}
    \vspace{-0.05in}
    \label{tbl:parameter}
    {\fontsize{7.8}{10}\selectfont
        \begin{tabular}{p{1.2cm}
            >{\centering\arraybackslash}m{1.2cm}
            >{\centering\arraybackslash}m{1.2cm}
            >{\centering\arraybackslash}m{1.2cm}
            >{\centering\arraybackslash}m{1.2cm}
        }
            \Xhline{3\arrayrulewidth}
            \multicolumn{5}{c}{\fontsize{7.8}{10}\selectfont{\bf Billon-scale ANN Datasets}} \\ 
            \Xhline{1.5\arrayrulewidth}
            & {\fontsize{7.3}{10}\selectfont{\textbf{SIFT}}} & {\fontsize{7.3}{10}\selectfont{\textbf{DEEP}}} & {\fontsize{7.3}{10}\selectfont{\textbf{Text2Image}}} & {\fontsize{7.3}{10}\selectfont{\textbf{MSSPACEV}}} \\
            \textbf{Data Type} & uint8 & fp32 & fp32 & int8 \\
            \textbf{Dimension} & 128 & 96 & 200 & 100 \\
            \Xhline{1.5\arrayrulewidth}
            \multicolumn{5}{c}{\fontsize{7.8}{10}\selectfont{\bf Search Parameters}} \\ 
            \Xhline{1.5\arrayrulewidth}
            \multicolumn{1}{c}{max\_degree} & \multicolumn{4}{l}{\quad Maximum number of neighbors per node} \\
            \multicolumn{1}{c}{cand\_list\_len} & \multicolumn{4}{l}{\quad Candidate list size} \\
            \multicolumn{1}{c}{num\_clusters} & \multicolumn{4}{l}{\quad Total number of clusters} \\
            \multicolumn{1}{c}{num\_probes} & \multicolumn{4}{l}{\quad Number of clusters searched per query} \\
            \Xhline{3\arrayrulewidth}
        \end{tabular}
    }
    \vspace{-0.06in}
\end{table}

\section{\Ours: Architecture and Mechanisms}

\subsection{System Architecture and Workflow}

\Ours utilizes a CXL-based architecture with a host CPU, CXL switch, and multiple CXL memory devices (Fig.~\ref{fig:overview}(a)).
Each CXL device consists of a CXL controller with a CXL-PNM module and DRAM devices supporting rank-level PUs.
The host dispatches queries via the switch to relevant CXL devices.
Each device performs local ANNS using its GPC (performing graph traversal and candidate list management) and returns local top-k results.
The host aggregates these for the global top-k.
Interface registers mapped in host memory facilitate host-PNM communication. Intermediate results generated during computation are stored in the temporary buffers, minimizing unnecessary memory access.

\Ours exploits DRAM rank-level parallelism for performance. Data is column-wise partitioned across ranks, allowing independent processing within each rank's PU, alleviating channel contention. Rank-level PUs compute partial distances (e.g., for L2 distance and inner product) on 64B sub-vector segments in parallel (Fig.~\ref{fig:overview}(c)).
Unlike prior work (CXL-ANNS~\cite{atc-2023-CXLANNS}), which offloaded only distance calculation to a domain-specific accelerator and required host-managed traversal, \Ours fully offloads traversal to the CXL GPC and uses rank-level PUs.
This significantly reduces PCIe traffic (only local top-k results are transferred) and memory bandwidth bottlenecks, enabling scalable ANNS for billion-scale data.

\subsection{Memory Space Management}

\Ours integrates CXL Host-managed Device Memory (HDM) into the host physical address (HPA) space using static mapping, eliminating runtime translation overhead. Following~\cite{atc-2022-directCXL}, the kernel driver maps HDM regions into HPA during enumeration and informs the devices. User applications allocate HDM via a namespace interface and \texttt{mmap()} system call. A segment table ensures contiguous physical/virtual mappings.  

Given ANNS's read-only nature after indexing, graphs and embedding data can be allocated using a static memory layout. This eliminates the need for dynamic virtual-to-physical address translation. During preprocessing, both the graph and embedding data are placed in HDM, and their metadata (e.g., base addresses and sizes) is registered with the controller.
Address calculation becomes simple arithmetic:

\vspace{-0.05in}
{\small
\begin{minipage}{\columnwidth}
\begin{align*}
& \text{addr}_{\text{node}} = \text{addr}_{\text{graph\_base}} + (\text{node\_index} \times \text{node\_stride}) \\
& \text{addr}_{\text{vector}} = \text{addr}_{\text{embedding\_base}} + (\text{vector\_index} \times \text{vector\_stride})
\end{align*}
\end{minipage}
}

\noindent To ensure mapping validity within the CXL device, the \texttt{mlock()} system call pins HDM regions in physical memory. 
This prevents any swapping or migration of the allocated memory, thereby maintaining a consistent address mapping.

\subsection{Adjacency-aware Cluster Placement}

\label{sec:Cluster-based}

\begin{algorithm}[tb!]
    \footnotesize  
    \caption{Adjacency-aware Cluster Placement}
    \label{algo:algorithm}
    \textbf{Input:} $cluster$: A cluster to be placed, including $.size$ and\\
    \hspace*{\algorithmicindent} ~~~~~~~~~~~~~~ a proximity-ordered list $.adj$ of the nearby clusters.\\
    \hspace*{\algorithmicindent} ~~~ $devices$: A list of available CXL devices in the system.\\
    \textbf{Output:} $best\_d$: The best CXL device for placing the $cluster$. \\

    \begin{algorithmic}[1]
        \State $best\_d, max\_cap, min\_loss \gets -1, 0, \infty$
        \For{$d$ in $devices$}
            \If{$d.remain \geq cluster.size$}
                \State $loss, proximity \gets 0, num\_devices$
                \For{$adj$ in $cluster.adj$}
                    \If{$adj \in d.clusters$}
                        \State $loss \gets loss + proximity$
                    \EndIf
                    \State $proximity \gets proximity - 1$
                \EndFor
                \If{($best\_d = -1$) \textbf{or} ($loss < min\_loss$) \textbf{or}
                      \Statex \hspace*{\algorithmicindent} \qquad \quad ($loss = min\_loss$  \textbf{and} $d.remain > max\_cap$) }
                    \State $best\_d, min\_loss, max\_cap \gets d, loss, d.remain$
                \EndIf
            \EndIf
        \EndFor 
        \State $best\_d.remain \gets best\_d.remain - cluster.size$
        \State \Return $best\_d$
    \end{algorithmic}
\end{algorithm}

Partitioning datasets into clusters for parallel search is common; however, it risks load imbalance if nearby clusters reside on the same CXL device. This issue is exacerbated when multiple queries target similar regions. We propose \emph{adjacency-aware cluster placement} (Algorithm~\ref{algo:algorithm}) to distribute adjacent clusters across different devices, enhancing parallelism.

All clusters are initially sorted by size in descending order, prioritizing the placement of larger clusters first. 
For each cluster, it calculates adjacency penalties (referred to as \emph{loss}) for devices with sufficient capacity (line~\circled{3}). Penalties increase based on the proximity of neighboring clusters already on a device (lines~\circled{5}$\sim$\circled{8}). The cluster is assigned to the device with the lowest penalty, the one with greater remaining capacity in case of ties (lines~\circled{9}$\sim$\circled{10}). As opposed to the CXL-ANNS's hop-count-based round-robin placement that ignores topology, our algorithm considers cluster adjacency.  

The host identifies k-nearest clusters via centroids and dispatches searches to the corresponding devices. With adjacent clusters distributed, traversals proceed in parallel, maximizing utilization.
In Section~\ref{sec:eval-algo}, we analyze the effect of our cluster placing algorithm.

\section{Evaluation}

\subsection{Experimental Setup}

To evaluate the performance of \Ours quantitatively, we developed a simulator integrated with Ramulator~\cite{ramulator}. 
Our setup models a 1TB CXL memory comprising four CXL devices, each with four DDR5-4800 channels and two ranks of 16Gb $\times$4 DRAM chips per channel (256GB per device).

We used two representative billion-scale datasets, SIFT1B and DEEP1B, from the BigANN benchmark~\cite{bigann}.
We incorporated a clustering mechanism into DiskANN (in-memory mode)~\cite{nips-2019-diskann} and extracted node visit traces from 10,000 queries per dataset to emulate realistic access patterns. 
These traces were used as input to our simulator to model the memory access patterns of the three main query processing operations: graph traversal, distance calculation, and candidate updates.

The generated memory requests were injected into Ramulator to measure query latency and analyze memory access behavior under various system configurations. We modeled a streaming scenario where queries are dispatched to the first available CXL device, enabling query-level parallelism.
To evaluate data placement, we compared our adjacency-aware algorithm with round-robin (RR) placement, which ignores inter-cluster proximity.

\begin{figure}[!t]
  \center
  \includegraphics[width=0.92\columnwidth]{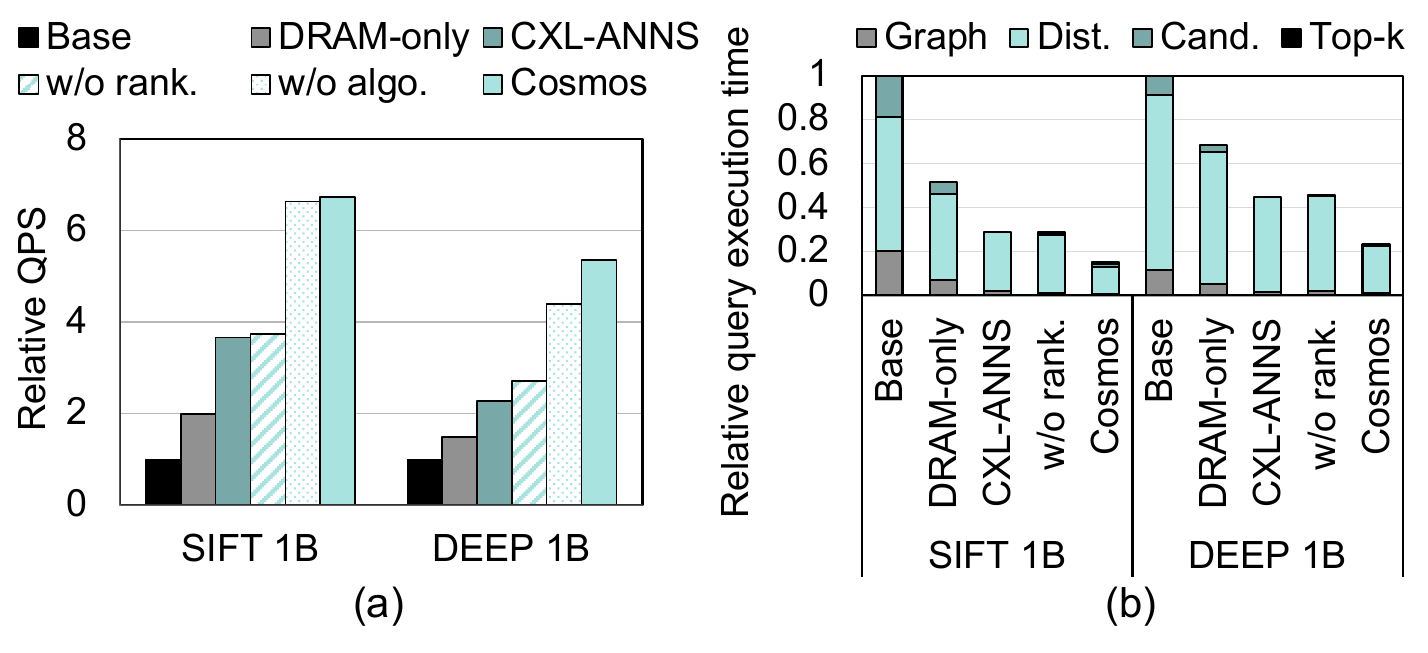}
  \vspace{-0.06in}
  \caption{(a) Relative query throughput (Query Per Second, QPS). (b) Breakdown of query execution time.}
  \label{fig:eval-perf}
   \vspace{-0.09in}
\end{figure}

\subsection{Overall Performance}

Fig.~\ref{fig:eval-perf}(a) illustrates the relative throughput (Queries Per Second, QPS) of various methods normalized to the \textbf{Base}, where all data resides in CXL-memory and computations are performed on the host side. The \textbf{DRAM-only} scenario assumes unlimited DRAM capacity, placing all data within DRAM. 
\textbf{CXL-ANNS}~\cite{atc-2023-CXLANNS} improves performance by offloading distance computation, applying fine-grained query scheduling, and hop-count-based graph caching; we reproduced the first two but excluded caching as it is beyond the scope of this work, which only affects graph traversal and has a negligible impact on total latency (Fig.~\ref{fig:eval-perf}(b)). To evaluate each component of \textbf{\Ours}, we evaluated three configurations: (1) without rank-level PU (\textbf{w/o rank.}), (2) without the data placement algorithm (\textbf{w/o algo.}), and (3) the full system (\textbf{\Ours}). 

\textbf{\Ours} achieves the highest performance, improving QPS by 6.72$\times$ (SIFT1B) and 5.35$\times$ (DEEP1B) over \textbf{Base}. While \textbf{DRAM-only} eliminates host-device data transfers, it is still bandwidth-limited. \textbf{CXL-ANNS} performs better by leveraging offloading and scheduling, but frequent transfers and bandwidth bottlenecks remain. \textbf{\Ours} addresses both issues by fully offloading graph traversal to CXL-side GPCs and accelerating distance computation using rank-level PUs.

Fig.~\ref{fig:eval-perf}(b) shows the single query latency breakdown within a single CXL device, excluding the impact of the data placement. \textbf{\Ours} significantly reduces graph traversal and distance calculation latency by combining in-memory execution with rank-level parallelism. \textbf{DRAM-only} benefits from reduced data movement, and \textbf{CXL-ANNS} reduces latency through scheduling, but neither entirely eliminates bandwidth-related overhead as \textbf{\Ours} does.

\subsection{Effectiveness of cluster placing}
\label{sec:eval-algo}

\begin{figure}[!t]
  \center
  \includegraphics[width=0.94\columnwidth]{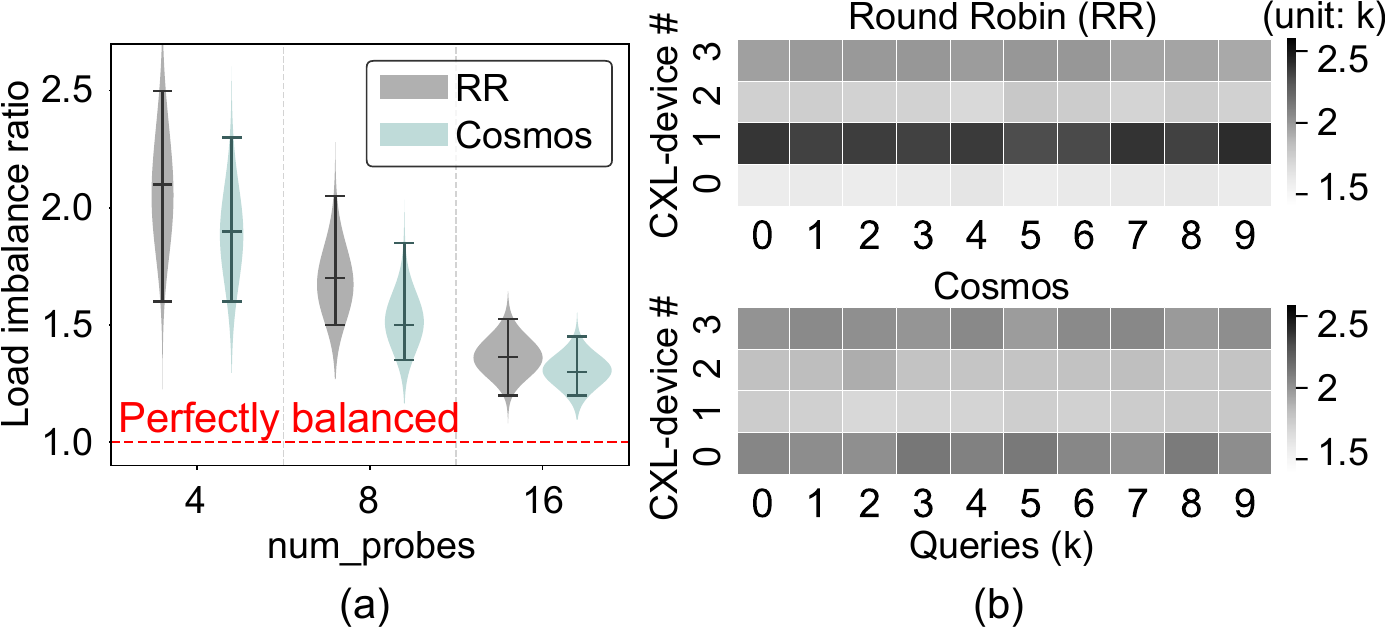}
  \vspace{-0.08in}
  \caption{(a) Load imbalance ratio according to the increasing number of probes. (b) Heatmap showing the number of clusters handled per device.}
  \label{fig:eval-algo}
   \vspace{-0.08in}
\end{figure}

Fig.~\ref{fig:eval-algo} highlights the effectiveness of our adjacency-aware data placement algorithm (Algorithm~\ref{algo:algorithm}). To isolate its impact, we fixed all other system configurations and compared against a baseline that distributes clusters across CXL devices in a round-robin (\textbf{RR}) manner. Fig.~\ref{fig:eval-algo}(a) shows the load imbalance ratio (LIR) across devices, defined as the maximum device load divided by the ideal uniform load under perfect distribution. Lower values indicate better load balancing. 

Across varying num\_probes (4, 8, and 16), \textbf{\Ours} effectively balances query load across devices, showing consistently lower LIR than \textbf{RR}. Fig.~\ref{fig:eval-algo}(b) presents a heatmap of cluster assignments handled per device over 10k queries. Unlike \textbf{RR}, which leads to uneven device utilization, \textbf{\Ours} ensures a uniform distribution. 
By relying solely on centroid distances and cluster sizes, without additional profiling, \textbf{\Ours} effectively balances query load across CXL devices, thereby enhancing system scalability and maximizing parallelism.

\section{Conclusion}
\label{sec:conclusion}

We have introduced \Ours, a scalable, full in-memory ANNS system designed to overcome the memory bandwidth and data movement bottlenecks inherent in billion-scale vector search.
By integrating programmable cores and rank-level processing units within CXL devices, \Ours eliminates host intervention during search and maximizes memory bandwidth utilization through parallel distance computation.
Further, we proposed an adjacency-aware data placement algorithm that effectively balances search load across CXL devices by strategically distributing neighboring clusters, enhancing parallelism and scalability.
Our evaluations demonstrated that \Ours significantly outperforms existing DRAM-based and prior CXL-based approaches in query throughput and latency.

\bibliographystyle{IEEEtran}
\bibliography{ref}

\end{document}